\documentclass[bibyear]{aa} 
\usepackage{epsfig}
\usepackage{amsmath}
\usepackage{subfigure}
\usepackage{natbib}
\usepackage{multirow}
\usepackage{color}
\usepackage{longtable}
\usepackage{graphicx}
\usepackage{epstopdf}
\usepackage{booktabs}
\usepackage{txfonts}

\newcommand{\jmst}{J.~Mol.~Struct.}   

\newcommand{\kms}{km\,s$^{-1}$}
\begin{document}

\title{Discovery of interstellar 3-cyano propargyl radical, CH$_2$CCCN \thanks{Based on observations carried out
with the Yebes 40m telescope (projects 19A003, 20A014, and 20D15). The 40m radio telescope at Yebes Observatory is operated by the Spanish Geographic Institute (IGN, Ministerio de Transportes, Movilidad y Agenda Urbana).}}

\author{
C.~Cabezas\inst{1},
M.~Ag\'undez\inst{1},
N.~Marcelino\inst{2,3},
B.~Tercero\inst{2,3},
J.~R.~Pardo\inst{1},
P.~de~Vicente\inst{3}
and
J.~Cernicharo\inst{1}
}

\institute{Grupo de Astrof\'isica Molecular, Instituto de F\'isica Fundamental (IFF-CSIC), C/ Serrano 121, 28006 Madrid, Spain.
\email carlos.cabezas@csic.es; jose.cernicharo@csic.es
\and Observatorio Astron\'omico Nacional (IGN), C/ Alfonso XII, 3, 28014, Madrid, Spain.
\and Centro de Desarrollos Tecnol\'ogicos, Observatorio de Yebes (IGN), 19141 Yebes, Guadalajara, Spain.
}

\date{Received; accepted}

\abstract{ We report the first detection in interstellar space of the 3-cyano propargyl radical (CH$_2$C$_3$N). This species was observed in the cold dark cloud TMC-1 using the Yebes 40m telescope. A total of seven rotational transitions for both ortho- and para-CH$_2$C$_3$N species were observed in the 31.0-50.4 GHz range. We derive a total column density of (1.6$\pm$0.4)$\times$10$^{11}$ cm$^{-2}$ and an ortho/para ratio of 2.4$\pm$1.2, which implies an abundance ratio CH$_2$C$_3$N/CH$_3$C$_3$N $\sim$ 0.1, in sharp contrast with the smaller analogues, in which case CH$_2$CN/CH$_3$CN $\sim$ 3. This indicates that the chemistry of the cyanides CH$_2$C$_3$N and CH$_3$C$_3$N behaves differently to that of the smaller analogues CH$_2$CN and CH$_3$CN. According to our chemical model calculations, the radical CH$_2$C$_3$N is mostly formed through the neutral-neutral reactions C + CH$_2$CHCN, C$_2$ + CH$_3$CN, and CN + CH$_2$CCH together with the dissociative recombination of the CH$_3$C$_3$NH$^+$ ion with electrons. The neutral-neutral reaction N + C$_4$H$_3$ could also lead to CH$_2$C$_3$N, although its role is highly uncertain. The identified radical CH$_2$C$_3$N could play a role in the synthesis of large organic N-bearing molecules, such as benzonitrile ($c$--C$_6$H$_5$CN) or nitrogen heterocycles.}


\titlerunning{Interstellar CH$_2$CCCN radical}
\authorrunning{Cabezas et al.}

\maketitle

\section{Introduction}

Cold dark clouds such as TMC-1 present a rich and complex chemistry that leads to the formation of a great variety of molecules. The list of molecular species observed in these clouds includes cations, mostly protonated forms of closed-shell abundant molecules (e.g. \citealt{Agundez2015, Marcelino2020,Cernicharo2020a,Cernicharo2021a,Cernicharo2021b}), and some hydrocarbon and nitrile anions (e.g. \citealt{Cernicharo2020b}), although most of the detected species are electrically neutral. Closed-shell species constitute only one-third of the neutral species observed in cold dark clouds, but they are the most abundant ones. The remaining two-thirds of the neutral species detected in these cold environments are open-shell radicals. Apart from OH, CH, C$_2$H, C$_4$H, C$_6$H, CH$_2$CCH, and NO, observed radicals have low abundances because, as ions, they are highly reactive species (see e.g. \citealt{Agundez2013}). Another fact that complicates the detection of radicals is the spectral dilution resulting from line splitting due to the interaction of the rotational  angular momentum with different types of angular momenta, such as the electronic orbital, the electron spin, or the nuclear spin.

Cyanomethyl radical, CH$_2$CN, is the simplest member of the CH$_2$C$_n$N ($n$$\eqslantgtr$1) radical series. It is derived from the closed-shell species CH$_3$CN by removing one hydrogen atom. CH$_2$CN was detected in the cold dark cloud TMC-1 by \cite{Saito1988} and \cite{Irvine1988}. Recent observations of TMC-1 using the QUIJOTE\footnote{\textbf{Q}-band \textbf{U}ltrasensitive \textbf{I}nspection \textbf{J}ourney to the \textbf{O}bscure \textbf{T}MC-1 \textbf{E}nvironment} line survey \citep{Cernicharo2021c} show that CH$_2$CN is a fairly abundant radical with a CH$_2$CN/CH$_3$CN ratio of 3.2$\pm$0.4. \citep{Cabezas2021a}. Hence, it is expected that larger members of the CH$_2$C$_n$N series can be observed in this source as well. CH$_2$CCN is the next member of the series. This radical is known as $\alpha$-cyanovinyl radical, and it has been characterized experimentally. Its rotational spectrum in the centimetre and millimetre regions has been observed \citep{Tang2000,Seiki2000,Prozument2013}, as has its Fourier-transform infrared emission spectrum \citep{Letendre2002}. However, the data available in the literature do not allow the precise predictions needed to search for it in the interstellar medium (ISM) to be obtained, and its astronomical detection has not been claimed so far.

The largest member of the CH$_2$C$_n$N radical series that has been characterized in the laboratory is the 3-cyano propargyl radical, CH$_2$C$_3$N (see Fig. \ref{mol_h2c4n}). \citet{Chen1998} detected the ortho-CH$_2$C$_3$N using Fourier transform microwave spectroscopy, and three years later \citet{Tang2001} improved the rotational parameters for CH$_2$C$_3$N by observing lines for the para-CH$_2$C$_3$N species using the same spectroscopy technique. \citet{Chen1998} also reported on their astronomical search for the CH$_2$C$_3$N radical, which they carried out using their experimental data. The obtained upper limit allowed them to estimate a column density of 2$\times$10$^{11}$cm$^{-2}$ in TMC-1.

In this letter we report the first identification of the CH$_2$C$_3$N radical in space towards TMC-1, based on the laboratory data previously reported by \citet{Chen1998} and \citet{Tang2001}. The derived column density for this radical is compared with analogue radicals and closed-shell species and is interpreted via chemical models to understand the chemical processes in which it is involved.

\section{Observations}

The data presented in this work are part of the QUIJOTE spectral line survey in the Q band towards TMC-1 ($\alpha_{J2000}=4^{\rm h} 41^{\rm  m} 41.9^{\rm s}$ and $\delta_{J2000}=+25^\circ 41' 27.0''$) that was performed at the Yebes 40m radio telescope during various observing sessions between November 2019 and April 2021. A total of 30 new molecular species have been detected using this survey \citep{Cernicharo2020a, Cernicharo2020b, Cernicharo2020c,  Marcelino2020, Marcelino2021, Agundez2021a, Agundez2021b, Cabezas2021a, Cabezas2021b, Cabezas2021c, Cernicharo2021a, Cernicharo2021b, Cernicharo2021c, Cernicharo2021d, Cernicharo2021e, Cernicharo2021f, Cernicharo2021g, Cernicharo2021h}. All observations were carried out using the frequency switching technique \citep{Cernicharo2019}, with a frequency throw of 10\,MHz during the two first observing runs and of 8\,MHz in the later ones. This observing mode provides a S/N that is $\sqrt{2}$ higher than the unfolded
data, but on the other hand it produces negative spectral features at $\pm$ 10\,MHz or 8\,MHz of each rotational transition. These negative features can be easily identified because of their symmetric displacement by exactly the frequency throw. As shown in Fig. \ref{spectra_oh2c4n}, we blanked these channels with negative features for the sake of convenience. The selected temperature scale is $T_A^*$. The $T_{MB}$ can easily be obtained by dividing the observed $T_A^*$ by the beam efficiency. Values of $\eta_{MB}$ have been provided by \citet{Tercero2021}. The $T_A^*$ was calibrated using two absorbers at different temperatures and the atmospheric transmission model ATM \citep{Cernicharo1985, Pardo2001}.

Different frequency coverages were observed, 31.08-49.52 GHz and 31.98-50.42 GHz, which permitted us to verify that no spurious ghosts were produced in the down-conversion chain. In this chain the signal coming from the receiver is down-converted to 1-19.5 GHz and then split into eight bands with a coverage of 2.5 GHz, each of which is analysed by the fast Fourier transform (FFT) spectrometers. Calibration uncertainties were adopted to be 10~\%, based on the observed repeatability of the line intensities between different observing runs. All data were analysed using the GILDAS package\footnote{\texttt{http://www.iram.fr/IRAMFR/GILDAS}}.

\section{Results}

\begin{figure}
\centering
\includegraphics[angle=0,width=0.4\textwidth]{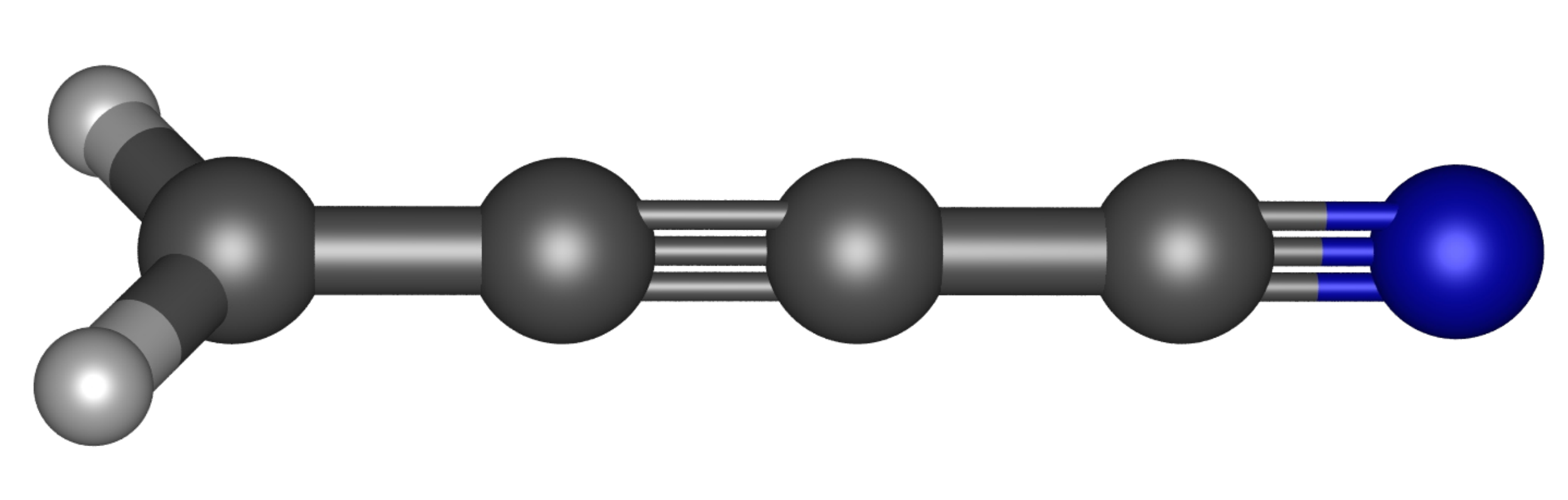}
\caption{Molecular structure of the CH$_2$C$_3$N radical.} \label{mol_h2c4n}
\end{figure}

\begin{figure}
\centering
\includegraphics[angle=0,width=0.47\textwidth]{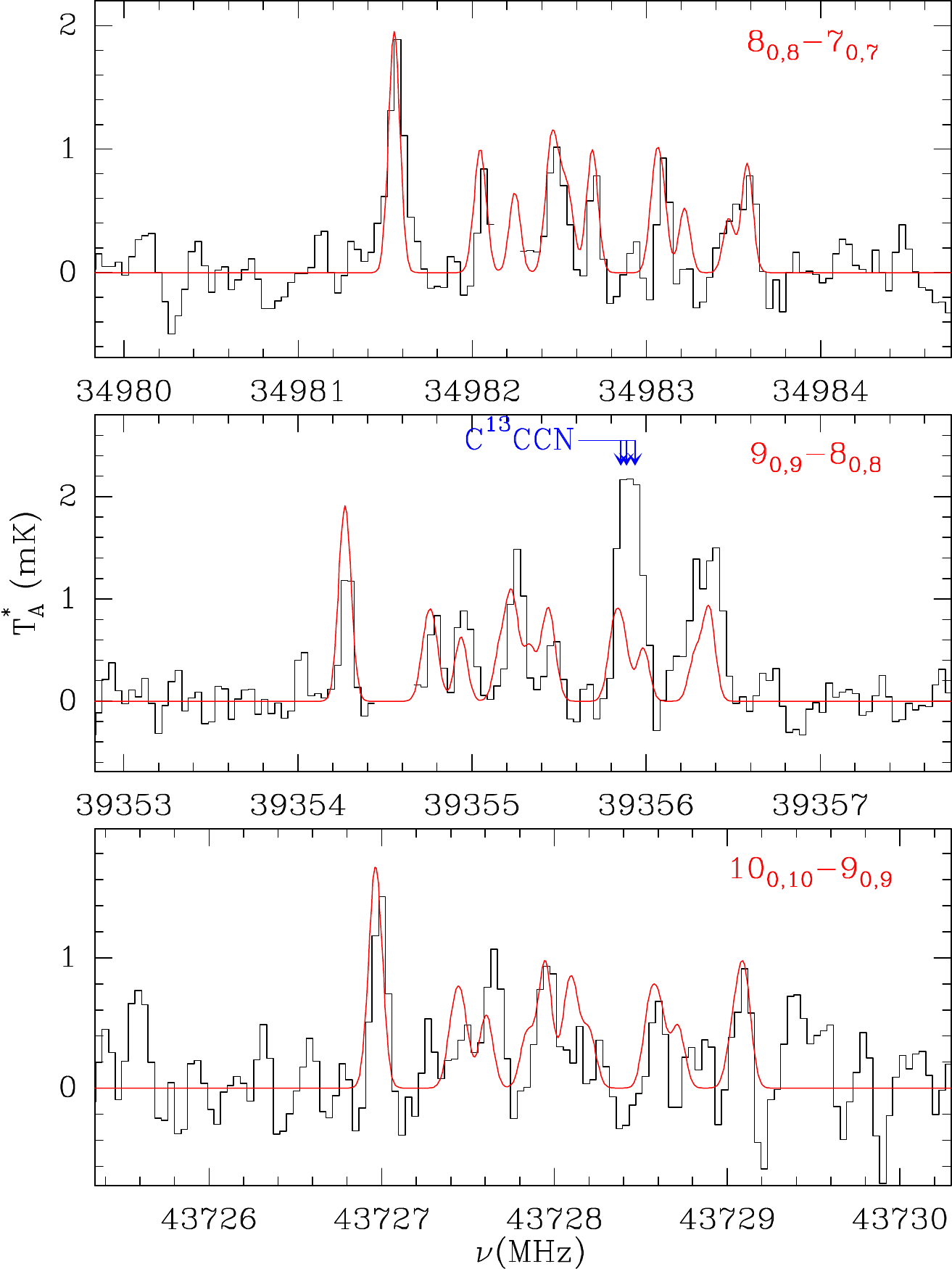}
\caption{Observed $N_{K_{a,}K_{c}}$= 8$_{0,8}$-7$_{0,7}$, 9$_{0,9}$-8$_{0,8}$, and 10$_{0,10}$-9$_{0,9}$ lines of ortho-CH$_2$C$_3$N in TMC-1 in the 31.0-50.4 GHz range. The most intense hyperfine components for each rotational transition are shown. The abscissa corresponds to the rest frequency, assuming a local standard of rest velocity of 5.83\,\kms. Blanked channels correspond to negative features produced in the folding of the frequency switching data. The ordinate is antenna temperature in millikelvins. Curves shown in red are the computed synthetic spectra.} \label{spectra_oh2c4n}
\end{figure}

\begin{figure*}
\centering
\includegraphics[angle=0,width=\textwidth]{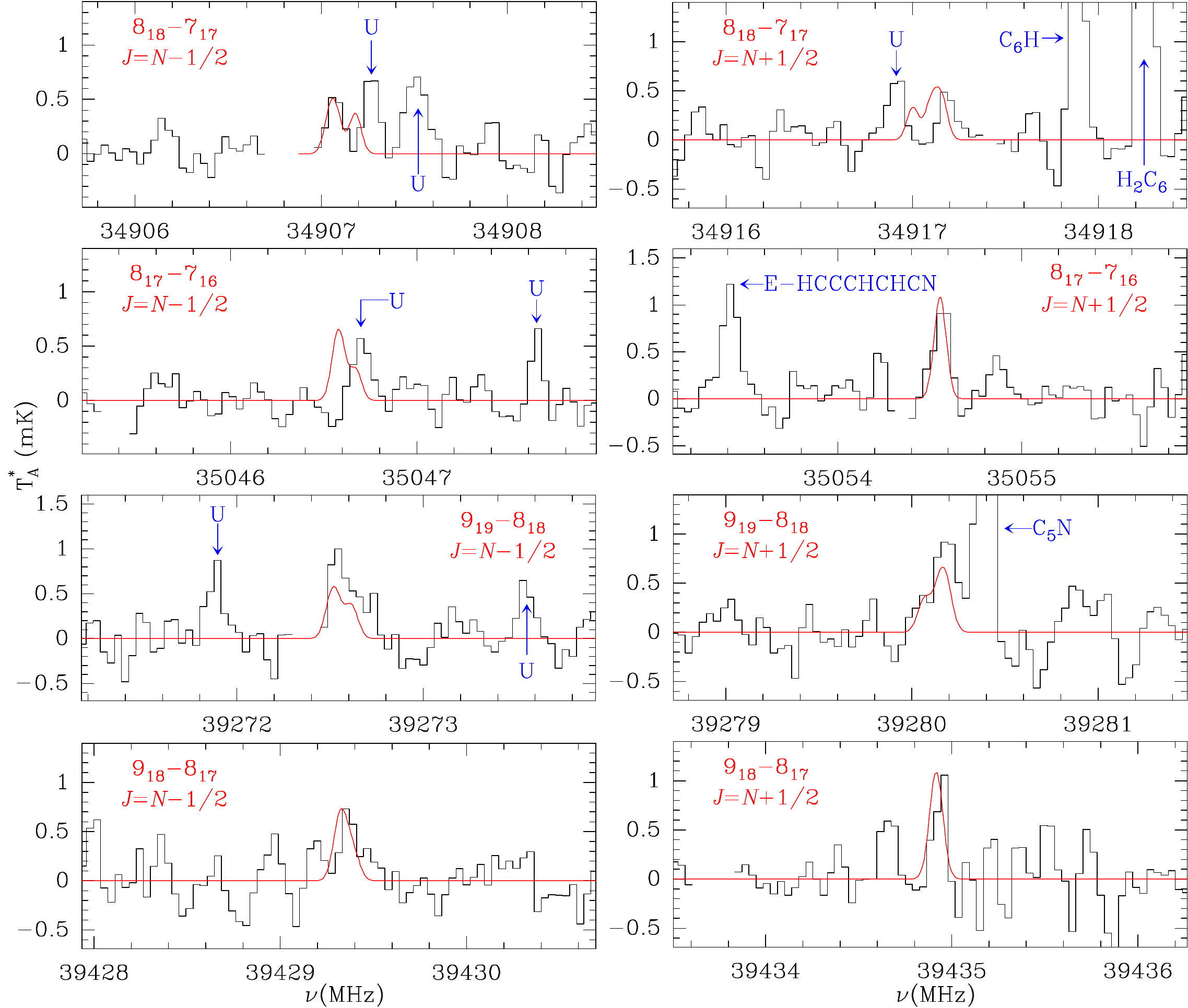}
\caption{Observed rotational transitions of para-CH$_2$C$_3$N in TMC-1 in the 31.0-50.4 GHz range. The most intense hyperfine components for each rotational transition are shown. The abscissa corresponds to the rest frequency, assuming a local standard of rest velocity of 5.83\,\kms. The ordinate is antenna temperature in millikelvins. Curves shown in red are the computed synthetic spectra.} \label{spectra_ph2c4n}
\end{figure*}

The 3-cyano propargyl radical, CH$_2$C$_3$N, is an asymmetric top molecule (see Fig. \ref{mol_h2c4n}) with a doublet electronic ground state ($^2B_1$) and a fairly large dipole moment of 4.43\,D \citep{Tang2001}. Due to its molecular symmetry, C$_{2v}$, it is necessary to discern between ortho-CH$_2$C$_3$N and para-CH$_2$C$_3$N levels, which are described by $K_a$ even and $K_a$ odd, respectively. \citet{Chen1998} observed in the laboratory four rotational transitions with $K_a$ = 0 for ortho-CH$_2$C$_3$N. A total of 89 hyperfine components were analysed with an effective Hamiltonian for a linear molecule in a $^2$$\Sigma$ electronic state, and a set of linear-molecule-like constants was determined. These constants can only be used to predict the rotational transitions for ortho-CH$_2$C$_3$N in a hypothetical astronomical search. \citet{Tang2001} confirmed the C$_{2v}$ structure for CH$_2$C$_3$N by measuring and analysing both $K_a$ = 1 for para-CH$_2$C$_3$N and those transitions previously measured by \citet{Chen1998} with $K_a$ = 0 for ortho-CH$_2$C$_3$N. In this manner, \citet{Tang2001} determined a total of 15 molecular constants for the asymmetric-rotor CH$_2$C$_3$N radical, which properly describe the rotational spectrum of this species and allow its radio-astronomical search, since both ortho- and para-CH$_2$C$_3$N are observable in the ISM.

Based only on the experimental data reported by \citet{Chen1998}, W. D. Langer \& T. Velusamy (cited as private communication in \citealt{Chen1998}) searched for ortho-CH$_2$C$_3$N in TMC-1. An upper limit of $T_A^*$ $\leqslant$5\,mK for the $N$= 5-4 transition at 21,864 MHz, averaged over the 0.20 km s$^{-1}$ spectral resolution, was obtained using the position switching observing mode with an on-source integration time of 6 hr. From these observations, W. D. Langer \& T. Velusamy estimate that the column density of CH$_2$C$_3$N in TMC-1 is $\leqslant$2$\times$10$^{11}$cm$^{-1}$ for an assumed dipole moment of 4.42 D, a rotational temperature of 10\,K, a line width of 0.5 km s$^{-1}$, and under the assumption that the source fills the telescope beam, 45$''$. Our QUIJOTE survey has an excellent sensitivity, with a $T_A^*$ rms noise level of 0.30\,mK per 38.15 kHz channel, which has allowed the detection of the 3-cyano propargyl radical.

Our search for CH$_2$C$_3$N is based on the frequency predictions for the ortho and para species made using the laboratory data from \citet{Chen1998} and \citet{Tang2001}. These predictions are available in the CDMS catalogue \citep{Muller2005}, entry number 064509, together with other data, such as partition functions. They were implemented in the MADEX code \citep{Cernicharo2012} to compute column densities. We considered the ortho and para species separately as there are no radiative or collisional transitions between them. The lowest energy level of the para species (1$_{1,1}$) is 13.9\,K above the ortho ground level (0$_{0,0}$). We adopted a dipole moment of 4.43\,D, as calculated by \citet{Tang2001}.

Four transitions of ortho-CH$_2$C$_3$N with $K_a$ = 0 are covered by our QUIJOTE survey, at 34.9, 39.3, 43.7, and 48.1 GHz. We observed three groups of lines at the predicted frequencies of the transitions 8$_{0,8}$-7$_{0,7}$, 9$_{0,9}$-8$_{0,8}$, and 10$_{0,10}$-9$_{0,9}$. Figure \ref{spectra_oh2c4n} shows the lines corresponding to these rotational transitions of ortho-CH$_2$C$_3$N with their hyperfine structure, as observed in TMC-1. Our model predicts an intensity of 1.0\,mK for the strongest hyperfine component of the 11$_{0,11}$-10$_{0,10}$ transition and between 0.4-0.7\,mK for the others. None of these lines are detected at the 3$\sigma$ (0.9\,mK) detection limit of the survey at this frequency. All the hyperfine components are detected with antenna temperatures between 0.6 and 1.9\,mK, and a few of them are blended with negative features produced in the folding of the frequency switching data (i.e. for the 8$_{0,8}$-7$_{0,7}$ transition). The hyperfine components of ortho-CH$_2$C$_3$N are precisely centred at the calculated frequencies with deviations in frequency smaller than 20 kHz, which is within the uncertainty given by the spectral resolution of 38.15 kHz and the error in the Gaussian fit.

A total of eight rotational transitions of para-CH$_2$C$_3$N with $K_a$ = 1 are covered by our survey. However, those with an $N$ quantum number higher than 9 are predicted at frequencies that are not detectable with the sensitivity of our survey. The spectral pattern of the $K_a$ = 1 transitions of para-CH$_2$C$_3$N is very different to that of the $K_a$ = 0 lines, where all the hyperfine components are spread over 2.0-2.5\,MHz. In contrast, each $K_a$ = 1 line is formed by two groups of hyperfine components separated by a few megahertz due to the spin-doubling interactions. This separation decreases with the $N$ quantum number. This is illustrated in Fig. \ref{spectra_ph2c4n}, where we show the four rotational transitions of para-CH$_2$C$_3$N with $K_a$ = 1 observed in our survey. Six groups of lines are clearly detected with a maximum antenna temperature of 0.6\,mK. The other two, 8$_{1,8}$-7$_{1,7}$ with $J$\,=\,$N$\,+\,1/2 and 8$_{1,8}$-7$_{1,6}$ with $J$\,=\,$N$\,$-$\,1/2, are affected by negative features produced in the folding of the frequency switching data and thus cannot be confirmed.

An analysis of the observed intensities using a line profile fitting method \citep{Cernicharo2021a} provides a rotational temperature of 7$\pm$1\,K, while the observed intensities are reproduced with column densities of (1.1$\pm$0.3)$\times$10$^{11}$ cm$^{-2}$ and (4.6$\pm$1.1)$\times$10$^{10}$ cm$^{-2}$ for ortho- and para-CH$_2$C$_3$N, respectively. The ortho/para ratio found is 2.4$\pm$1.2. \citet{Chen1998} reported an estimated column density for CH$_2$C$_3$N in TMC-1 $\leqslant$2$\times$10$^{11}$cm$^{-1}$, which is in accordance with our value for the total (ortho plus para) column density for CH$_2$C$_3$N of (1.6$\pm$0.4)$\times$10$^{11}$ cm$^{-2}$.

\section{Chemical modelling}
\label{sec:chemistry}

To examine the chemical processes that could form the radical CH$_2$C$_3$N in TMC-1, we carried out chemical modelling calculations. We adopted typical conditions of cold dark clouds: a gas kinetic temperature of 10 K, a volume density of H nuclei of 2\,$\times$\,10$^4$ cm$^{-3}$, a cosmic-ray ionization rate of H$_2$ of 1.3\,$\times$\,10$^{-17}$ s$^{-1}$, a visual extinction of 30 mag, and low-metal elemental abundances (see e.g. \citealt{Agundez2013}). We used the chemical network RATE12 from the UMIST database \citep{McElroy2013}, with updates from \cite{Loison2014} and \cite{Marcelino2021} to include C$_4$H$_3$N isomers.

Since the radical CH$_2$C$_3$N is not included in the UMIST database, we added some reactions to account for its formation and destruction. We assumed that CH$_2$C$_3$N is destroyed through reactions with the neutral atoms C, N, and O, as well as through reactions with the atomic cations C$^+$ and H$^+$, with rate coefficients similar to those involving the radical CH$_2$CN. As formation routes, we included the reactions\begin{equation}
\rm C + CH_2CHCN \rightarrow \rm CH_2C_3N + H, \label{reac:c+ch2chcn}
\end{equation}
\begin{equation}
\rm C_2 + CH_3CN \rightarrow \rm CH_2C_3N + H, \label{reac:c2+ch3cn}
\end{equation}
\begin{equation}
\rm CN + CH_2CCH \rightarrow \rm CH_2C_3N + H, \label{reac:cn+ch2cch}
\end{equation}
\begin{equation}
\rm CH_3 + C_3N \rightarrow \rm CH_2C_3N + H. \label{reac:ch3+c3n}
\end{equation}
Reaction~(\ref{reac:c+ch2chcn}) has been studied using crossed molecular beam experiments and theoretical calculations \citep{Su2005,Guo2006a}. These studies indicate that the reaction is barrier-less and occurs through H atom elimination, yielding as main products the radicals 1-cyano propargyl (HCCCHCN) and 3-cyano propargyl (CH$_2$C$_3$N). The latter is inferred to be produced at least twice less efficiently than the former \citep{Guo2006a}, and we thus adopted a rate coefficient of 10$^{-10}$ cm$^3$ s$^{-1}$ for reaction~(\ref{reac:c+ch2chcn}). Reaction~(\ref{reac:c2+ch3cn}) has not been studied to our knowledge, but some information can be extracted from the known reactivity of C$_2$ with unsaturated hydrocarbons, in particular with CH$_3$CCH. The reaction C$_2$ + CH$_3$CCH has been measured to be rapid, with a nearly constant rate coefficient of (4-5)\,$\times$10$^{-10}$ cm$^3$ s$^{-1}$ in the temperature range 77-296 K \citep{Daugey2008}. Moreover, crossed molecular beam experiments and theoretical calculations indicate that the radical CH$_2$C$_4$H is the preferred product, with an estimated branching ratio of 0.65 \citep{Guo2006b,Mebel2006}. Based on the behaviour of the reaction C$_2$ + CH$_3$CCH, we adopted a rate coefficient of 3\,$\times$\,10$^{-10}$ cm$^3$ s$^{-1}$ for reaction~(\ref{reac:c2+ch3cn}). The radical-radical reactions~(\ref{reac:cn+ch2cch}) and (\ref{reac:ch3+c3n}) have not been studied to our knowledge, although based on the known reactivity of CN and C$_3$N with unsaturated closed-shell hydrocarbons, it is plausible that they occur with no barrier, yielding CH$_2$C$_3$N as one of the main products. We thus adopted a rate coefficient of 10$^{-10}$ cm$^3$ s$^{-1}$ for them.

\begin{figure}
\centering
\includegraphics[angle=0,width=\columnwidth]{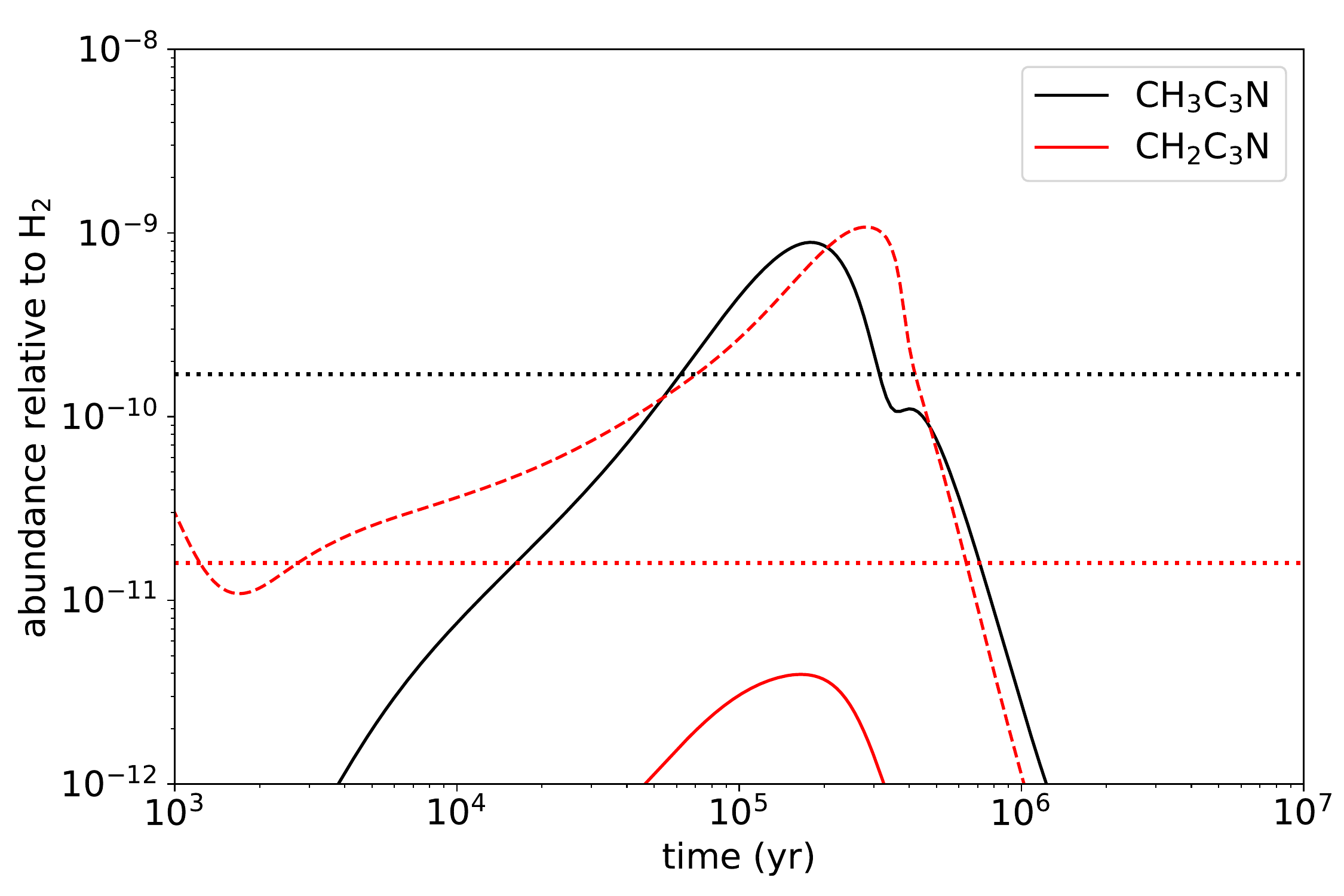}
\caption{Calculated fractional abundances of CH$_3$C$_3$N and CH$_2$C$_3$N as a function of time. The solid red line corresponds to the abundance of CH$_2$C$_3$N when the reaction N + C$_4$H$_3$ is neglected and the dashed red line to the case in which this reaction is included. Horizontal dotted lines correspond to the abundances observed in TMC-1.} \label{fig:abun}
\end{figure}

We also included an additional formation route to CH$_2$C$_3$N through the dissociative recombination of the precursor ion CH$_3$C$_3$NH$^+$ with electrons. \cite{Loison2014} estimated a branching ratio of 0.24 for the production of the radical CH$_2$CN in the dissociative recombination of the smaller analogue ion CH$_3$CNH$^+$, and we thus assumed that the same behaviour holds for the formation of the radical CH$_2$C$_3$N in the dissociative recombination of CH$_3$C$_3$NH$^+$. The reaction channel of formation of CH$_2$C$_3$N then reads
\begin{equation}
\rm CH_3C_3NH^+ + e^- \rightarrow \rm CH_2C_3N + H + H, \label{reac:ch3c3nh+}
\end{equation}
with a rate coefficient of 8\,$\times$\,10$^{-8}$ ($T$/300\,K)$^{-0.5}$ cm$^3$ s$^{-1}$.

In Fig.~\ref{fig:abun} we show, as a solid red line, the calculated fractional abundance of C$_2$H$_3$CN as a function of time. The peak abundance is about four times lower than the value observed in TMC-1. The main formation routes to CH$_2$C$_3$N are reactions~(\ref{reac:c+ch2chcn}), (\ref{reac:c2+ch3cn}), (\ref{reac:cn+ch2cch}), and (\ref{reac:ch3c3nh+}), while reaction~(\ref{reac:ch3+c3n}) is only a minor route.

There is another potential formation route to CH$_2$C$_3$N. In the case of the smaller analogue radical CH$_2$CN, the chemical model indicates that it is mainly formed through the dissociative recombination of CH$_3$CNH$^+$ with electrons, but also by the reaction N + C$_2$H$_3$, which is known to produce CH$_2$CN as a main product \citep{Payne1996}. Similarly, it is plausible that the reaction N + C$_4$H$_3$ yields CH$_2$C$_3$N. If this reaction is implemented in the chemical model with a rate coefficient similar to that of N + C$_2$H$_3$, then it becomes the main route to CH$_2$C$_3$N and its calculated peak abundance increases by more than two orders of magnitude, lying well above the observed value (see the dashed red line in Fig.~\ref{fig:abun}). However, it is currently not known whether the reaction N + C$_4$H$_3$ can produce CH$_2$C$_3$N. Moreover, the situation becomes more complicated because there are various possible isomers of the radical C$_4$H$_3$, although the chemical network does not distinguish between them. A dedicated study of the reaction N + C$_4$H$_3$ is needed to shed light on this point.

In Fig.~\ref{fig:abun} we also show the calculated abundance of the related molecule CH$_3$C$_3$N, which has a peak value about five times above the abundance observed in TMC-1. This behaviour is similar to that reported previously in \cite{Marcelino2021}.

\section{Discussion}
\label{sec:discussion}

In light of the discovery of the radical CH$_2$C$_3$N in TMC-1, it is worth comparing its abundance with that of chemically related molecules. In TMC-1 we have $N$(CH$_2$CN) = 1.5\,$\times$\,10$^{13}$ cm$^{-2}$ \citep{Cabezas2021a}, $N$(CH$_3$CN) = 4.7\,$\times$\,10$^{12}$ cm$^{-2}$ \citep{Cabezas2021a}, and $N$(CH$_3$C$_3$N) = 1.7\,$\times$\,10$^{12}$ cm$^{-2}$ \citep{Marcelino2021}. Therefore, the abundance ratio CH$_2$C$_3$N/CH$_3$C$_3$N is just 0.09, well below unity.\ This is in contrast with the smaller analogues, in which case the abundance ratio CH$_2$CN/CH$_3$CN is above one, concretely 3.2. This indicates that the chemistry of the cyanides CH$_2$C$_3$N and CH$_3$C$_3$N behaves differently to that of the smaller analogues CH$_2$CN and CH$_3$CN. Moreover, it is unclear whether the radical and the corresponding closed-shell molecule are connected by some common formation routes or have completely disconnected chemistries.

A common route to both CH$_2$CN and CH$_3$CN is the dissociative recombination of CH$_3$CNH$^+$. However, the yield ratio CH$_2$CN/CH$_3$CN is currently unconstrained (see \citealt{Vigren2008}). It is likely that this reaction is a major formation pathway to CH$_3$CN but not to CH$_2$CN. Even if some CH$_2$CN is formed during the dissociative recombination of CH$_3$CNH$^+$, most CH$_2$CN should be formed in TMC-1 through an independent and very efficient process, such as the reaction N + C$_2$H$_3$. It should be noted that this pathway must be much more efficient than the formation of CH$_3$CN through CH$_3$CNH$^+$ + e$^-$ to account for the higher abundance of CH$_2$CN compared to CH$_3$CN and the fact that CH$_2$CN should be far more reactive than CH$_3$CN. If true, this implies that the radical C$_2$H$_3$ should be abundant in TMC-1.

In the case of the larger cyanides CH$_2$C$_3$N and CH$_3$C$_3$N, a similar highly efficient route to the radical CH$_2$C$_3$N must be prevented since in this case the radical is substantially less abundant than the closed-shell molecule. This takes us to the reaction N + C$_4$H$_3$, which, according to our chemical model calculations, should not be a major source of CH$_2$C$_3$N, unlike in the case of the smaller analogue, where the reaction N + C$_2$H$_3$  is a major route to CH$_2$CN. In summary, the routes to the radicals CH$_2$CN and CH$_2$C$_3$N are likely chemically different.

A further point to discuss is whether the radical detected in this work could be an important intermediate to forming larger molecules. Although dedicated studies on the reactivity of CH$_2$C$_3$N are needed, our radical could play a role in the synthesis of large organic N-bearing molecules, such as nitrogen heterocycles. For example, benzonitrile ($c$-C$_6$H$_5$CN), known to be present in TMC-1 and other molecular clouds \citep{McGuire2018,Burkhardt2021} and thought to be formed through the reaction of benzene with CN \citep{Cooke2020}, could be formed through the reaction of CH$_2$C$_3$N with allene (CH$_2$CCH$_2$), which is suspected to be abundant in TMC-1 \citep{Marcelino2021,Cernicharo2021e,Agundez2021a}.

\section{Conclusions}

We have reported the detection of the 3-cyano propargyl radical (CH$_2$C$_3$N) in the cold dark cloud TMC-1. A total of seven rotational transitions with several hyperfine components were observed in our Q-band TMC-1 survey. This radical is about ten times less abundant than the corresponding closed-shell molecule CH$_3$C$_3$N, in contrast to the case of the smaller analogues, where the radical CH$_2$CN is three times more abundant than CH$_3$CN. Comparison between the chemical routes of the radicals CH$_2$CN and CH$_2$C$_3$N and the corresponding closed-shell species indicates that the formation routes to the radicals CH$_2$CN and CH$_2$C$_3$N are completely dissimilar.

\begin{acknowledgements}

This research has been funded by ERC through grant ERC-2013-Syg-610256-NANOCOSMOS. Authors also thank Ministerio de Ciencia e Innovaci\'on for funding support through projects PID2019-106235GB-I00 and PID2019-107115GB-C21 / AEI / 10.13039/501100011033. MA thanks Ministerio de Ciencia e Innovaci\'on for grant RyC-2014-16277.

\end{acknowledgements}

\end{document}